\documentclass[conference]{IEEEtran}
\IEEEoverridecommandlockouts
\usepackage{cite}
\usepackage{amsmath,amssymb,amsfonts}
\usepackage{algorithmic}
\usepackage{graphicx}
\usepackage{textcomp}
\usepackage{xcolor}
\usepackage{booktabs}

\usepackage[final]{changes}
\def\BibTeX{{\rm B\kern-.05em{\sc i\kern-.025em b}\kern-.08em
    T\kern-.1667em\lower.7ex\hbox{E}\kern-.125emX}}
\usepackage{todonotes}
\usepackage[hidelinks]{hyperref}

\begin{document}

%\title{Feasibility of simultaneous EEG-fMRI at 0.55 T: Recording and Denoising\\

\title{Feasibility of Simultaneous EEG-fMRI at 0.55 T: Recording, Denoising, and Functional  Mapping\\

\thanks{The authors receive grant support from the National Institutes of Health (R01EB026299, U01EB023820) and the National Science Foundation (No. 1828736).}
}

\author{
    \IEEEauthorblockN{
        Parsa Razmara\textsuperscript{1}, 
        Takfarinas Medani\textsuperscript{1},  
        Majid Abbasi Sisara\textsuperscript{2}, 
        Anand A. Joshi\textsuperscript{1}, 
        Rui Chen\textsuperscript{1}, \\
        Woojae Jeong\textsuperscript{1}, 
        Ye Tian\textsuperscript{1}, 
        Krishna S. Nayak\textsuperscript{1}, 
        and Richard M. Leahy\textsuperscript{1}
    }
    \vspace{0.15cm}
    \IEEEauthorblockA{
        \textsuperscript{1}\textit{Ming Hsieh Department of Electrical and Computer Engineering, University of Southern California}, Los Angeles, CA, USA \\
        \textsuperscript{2}\textit{Division of Biokinesiology and Physical Therapy, University of Southern California}, Los Angeles, CA, USA
    }
}

\maketitle

\begin{abstract}
Simultaneous recording of electroencephalography (EEG) and functional MRI (fMRI) can provide a more complete view of brain function by merging high temporal and spatial resolutions. This proof-of-concept study presents initial evidence for the feasibility of simultaneous EEG-fMRI at 0.55T in a visual task. We characterize the gradient and ballistocardiogram (BCG) artifacts inherent to this environment and demonstrate that the lower field strength suggests a reduction in the magnitude of the BCG artifact compared to high-field (1.5T, 3T, 7T) systems. This reduction shows promise for facilitating effective denoising while preserving signal integrity. Furthermore, we tested a multimodal integration pipeline that uses the EEG power envelope to compute a predictor of the hemodynamic BOLD response, demonstrating the potential for EEG-based estimation of neurovascular coupling in this environment. We demonstrate that combined EEG-fMRI at 0.55T is feasible and represents a promising environment for multimodal neuroimaging.

%Simultaneous recording of electroencephalography (EEG) and functional MRI (fMRI) provides a comprehensive view of brain function by merging high temporal and spatial resolutions. While high-field (3T+) systems are standard, they introduce severe technical trade-offs, including massive magnetic field-induced artifacts in the EEG signal and reduced compatibility with medical implants. This study investigates the feasibility of simultaneous EEG-fMRI at 0.55T using a high-performance gradient system. We demonstrate that task-based fMRI at 0.55T, utilizing 10-minute concatenated runs, yields robust activations in visual cortex comparable to 3T results (mean t-scores of 12.67 and 8.0, respectively). Furthermore, we establish that the lower magnetic field strength significantly reduces the ballistocardiogram (BCG) artifact, facilitating more effective EEG denoising. Our results validate 0.55T scanners as an advantageous platform for multimodal neuroimaging.
\end{abstract}


\section{Introduction}
Simultaneous electroencephalography (EEG) and functional MRI (fMRI) recording is a valuable multimodal approach that provides complementary insights into brain function, with EEG offering high temporal resolution and fMRI providing spatial detail \cite{b1, b2}. fMRI has traditionally relied on higher-field MRI systems ($\geq$3T) to maximize the blood oxygenation level-dependent (BOLD) sensitivity, resolution, and spatial specificity inherent at stronger magnetic fields \cite{b3, b4, b5, b6}. However, high fields also create a challenging environment for multimodal integration because of magnetic field-induced artifacts in the EEG signal \cite{b7}. Lower field scanners have the added advantage that studies can more readily be performed in subjects with implants including neurostimulators \cite{b8}.

%Simultaneous electroencephalography (EEG) and functional MRI (fMRI) recording is a valuable multimodal approach that provides complementary insights into brain function, with EEG offering high temporal resolution and fMRI providing spatial detail \cite{b9, b15}. EEG-fMRI studies have traditionally been conducted in high-field MRI environments \cite{b9}. Functional MRI itself has become a cornerstone of noninvasive human brain mapping since its introduction over three decades ago \cite{b1}. While historical studies utilized lower-field systems ($\leq$1T), it was long established that blood oxygenation level-dependent (BOLD) contrast generally improves with higher magnetic field strengths ($\geq$3T) due to increased sensitivity and resolution \cite{b2, b3, b5}. However, these benefits come with substantial trade-offs. High-field systems suffer from increased sensitivity to $B_0$ field inhomogeneities, reduced subject compatibility for those with neurostimulators or other implants, and increasingly severe artifacts when combined with complementary functional imaging modalities such as EEG \cite{b6, b7, b8}.

The recent resurgence of lower-field systems ($\leq$1T) for general MRI applications \cite{b8, b10, b11
, b12, b13, b14, b15}, driven by modern 0.55T systems equipped with high-performance gradients, offers a compelling alternative to high-field scanners for simultaneous EEG and fMRI \cite{b7, b8}. Unlike earlier lower-field scanners, these contemporary systems deliver superior imaging performance while offering distinct physical advantages for simultaneous EEG-fMRI. However, imaging at 0.55T operates with reduced intrinsic SNR, where signal estimation can benefit from statistical modeling and regularization principles in MRI analysis \cite{b24, razmara2024advancements}. The primary benefit lies in the reduced interaction between the static magnetic field and the EEG recording equipment. The lower field strength is expected to significantly attenuate the ballistocardiogram (BCG) artifact, a persistent noise source linked to cardiac motion in the magnetic field \cite{b7, b8}. This combination of high-performance imaging and reduced electromagnetic interference positions the 0.55T system as a potentially advantageous platform for simultaneous EEG-fMRI.%, while also expanding neuroimaging to populations with implanted devices who are currently excluded from high-field studies. 
%These characteristics position high-performance 0.55T systems as a unique uniquely advantageous platform for expanding neuroimaging to populations currently excluded from higher-field studies.

%Contemporary 0.55T MRI systems offer a more accessible alternative, broadening neuroimaging options in settings where high-field MRI is impractical or unavailable \cite{b14}. Lower $B_0$ field strengths provide several advantages, such as improved comfort (due to a larger bore), reduced susceptibility artifacts, lower specific absorption rate (SAR), decreased acoustic noise, and improved device compatibility \cite{b6}. 

Recent studies have demonstrated the feasibility of fMRI at 0.55T \cite{b16, b17, b18}. Nonetheless, integrating EEG with 0.55T MRI presents specific technical challenges, primarily due to gradient-induced artifacts (GA) and ballistocardiogram (BCG) interference in the EEG \cite{b7, b19}. These artifacts are present at all magnetic fields although the BCG is reduced at lower field strengths. They will exhibit unique characteristics at 0.55T, necessitating tailored denoising approaches to ensure reliable EEG signal quality. 

This pilot study investigates the feasibility of simultaneous EEG-fMRI recording within a 0.55T MRI environment, aiming to understand the impact of this environment on EEG data quality and to assess preliminary methods for reducing artifact interference. To our knowledge, this is the first study to investigate simultaneous EEG-fMRI recording at 0.55T. This study seeks to evaluate the potential of such a setup for neuroimaging applications and to develop foundational techniques that could enable future multimodal studies in diverse clinical and research environments. To compensate for the reduced intrinsic BOLD sensitivity of the 0.55T MRI signal, we used extended acquisition periods (10-minute concatenated runs). Our preliminary work at 0.55T \cite{b17} has demonstrated that by extending acquisition times, we can achieve substantial gains in sensitivity, yielding activation strengths comparable to standard high-field acquisitions \cite{b2, b20}. 

We describe and evaluate EEG-denoising methods. We then describe a multimodal integration pipeline to explore use of the EEG power envelope as a predictor of the hemodynamic BOLD response, to investigate the potential to model neurovascular coupling in this environment. Our results demonstrate that combined EEG-fMRI at 0.55T is feasible and represents a promising environment for multimodal neuroimaging.

The main contributions of this work are: (i) the first demonstration of simultaneous EEG-fMRI at 0.55T; (ii) quantitative characterization of gradient and BCG artifacts at this field strength, with comparison to published high-field benchmarks (1.5T, 3T, 7T); (iii) an EEG denoising pipeline that recovers spontaneous alpha rhythms and SSVEP harmonics in the scanner environment; and (iv) an EEG-informed BOLD regressor that produces voxel-wise activation maps consistent with a standard block-design analysis.

\section{Methods}

\subsection{Experimental Methods}
\added[id=P]{The experimental procedures involving human subjects described in this paper were approved by the Institutional Review Board of the University of Southern California.} As a pilot feasibility study, two healthy adult subjects were enrolled \deleted[id=P]{under a protocol approved by our Institutional Review Board, }after providing written informed consent. Data collection was performed under three distinct conditions to isolate artifact sources: 1) \textbf{Outside MRI}, where baseline EEG was recorded to capture true neural activity without scanner-induced interference; 2) \textbf{Inside MRI (Scanner OFF)}, capturing the contribution of the BCG—induced by cardiac motion within the static magnetic field—without GA interference; and 3) \textbf{Inside MRI (Scanner ON)}, involving fully simultaneous EEG-fMRI recording to capture the integrated effect of BCG and GA.

\begin{figure}[t]
\centerline{\includegraphics[width=\columnwidth]{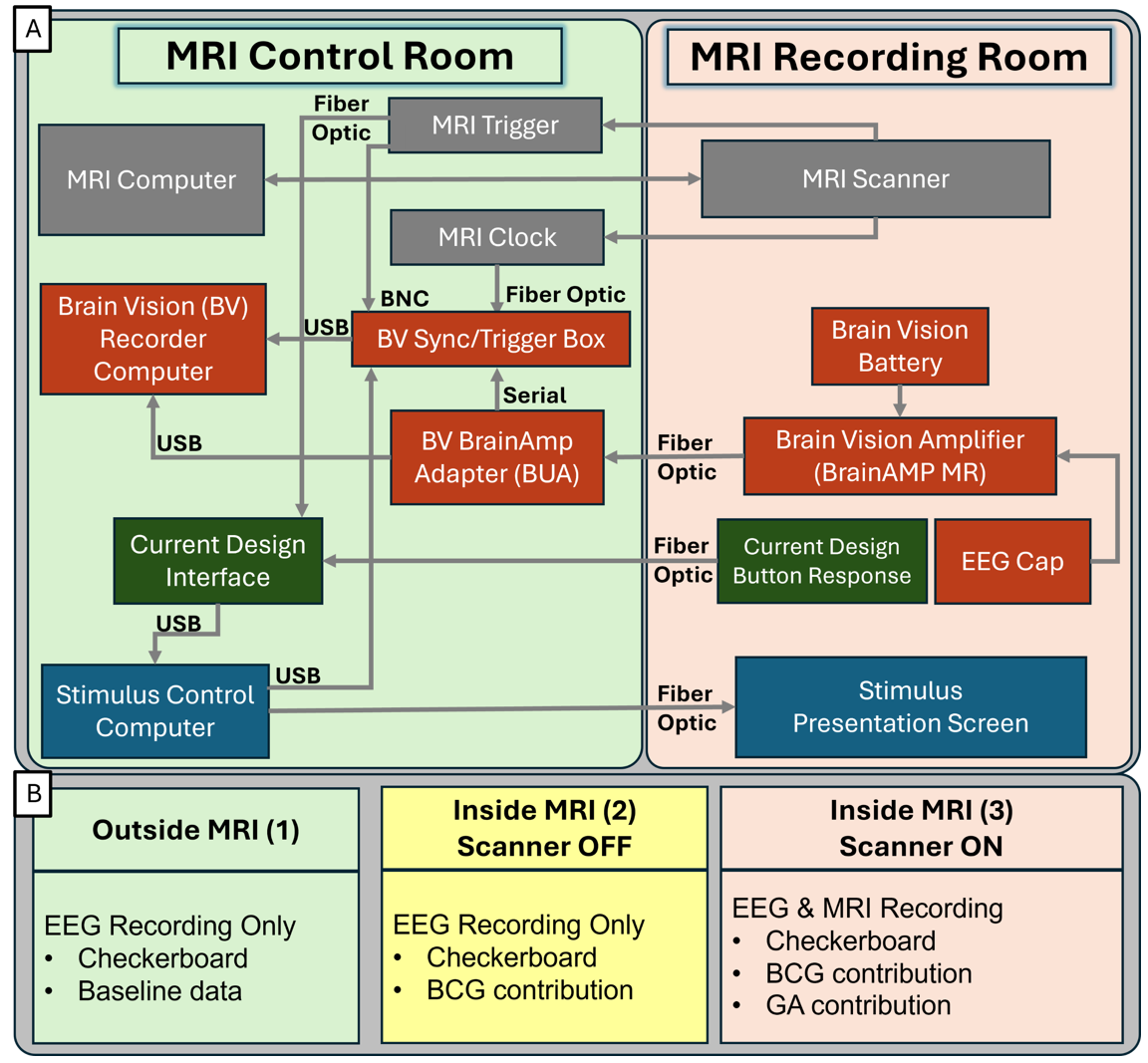}}
%\centerline{\includegraphics[width=\columnwidth]{1.tif}}
\caption{(A) Schematic of the simultaneous EEG-fMRI setup showing synchronization between the MRI trigger and the BrainVision recorder. (B) Protocol overview for Outside, Scanner OFF, and Scanner ON conditions. %\color{blue}{(Note: This figure will be updated with higher image quality.)}
}
\label{fig1}
\end{figure}

Tasks included a flickering checkerboard (visual stimulation) following an established block design. The visual task involved 24-second blocks alternating with rest \cite{b2, b17}. The experimental setup and protocol timing are illustrated in Fig. \ref{fig1}.

\subsection{Hardware and Data Acquisition}
Imaging was conducted on a 0.55T scanner (prototype MAGNETOM Aera, Siemens Healthineers, Forchheim, Germany) equipped with high-performance gradients (45 mT/m amplitude, 150 T/m/s slew rate) \cite{b8, b21} and a 16-channel head-neck receive array. Functional data were acquired using a single-shot EPI BOLD sequence (FOV: $210 \times 170$ mm$^2$, voxel size: $3.3 \times 3.3 \times 4$ mm$^3$, TR: 3000 ms, TE: 85 ms), with twenty interleaved axial slices to achieve whole-cerebrum coverage \cite{b17}. Simultaneous EEG was recorded using a 32-channel MR-compatible BrainCap and BrainAmp MR system (Brain Products GmbH, Gilching, Germany). To minimize GA jitter, the EEG system was phase-locked to the MRI clock via a sync-trigger box. GA correction and BCG removal were performed as described below. 
%in Fig. \ref{fig2}.
\begin{figure}[t]
\centerline{\includegraphics[width=\columnwidth]{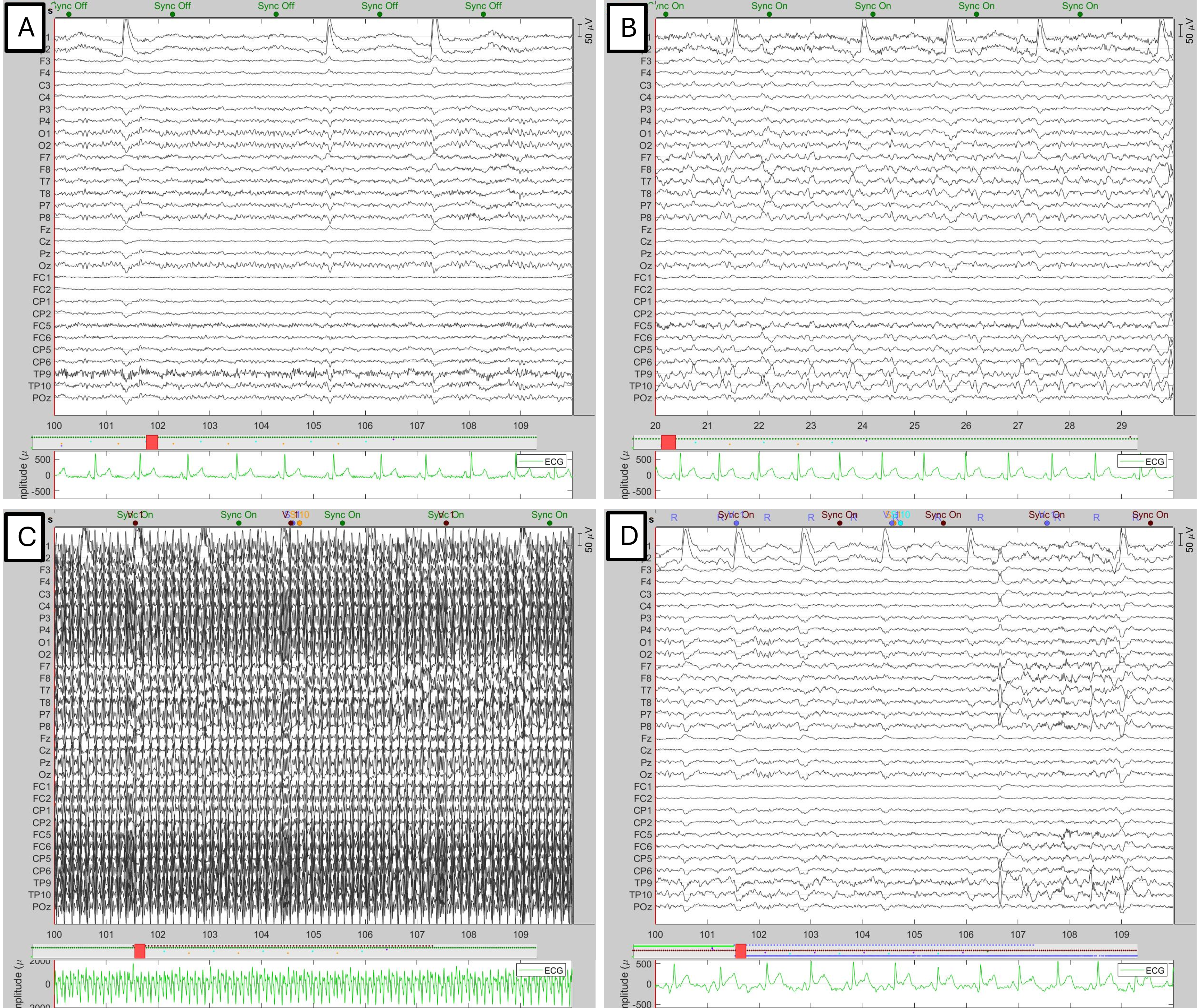}}
\caption{Representative example of EEG signal showing the raw signal across conditions: (A) Outside scanner, (B) Scanner OFF, (C) Scanner ON, and (D) Final cleaned data after artifact subtraction. %\color{blue}{(Note: This figure will be updated with higher image quality.)}
}\label{fig2}
\end{figure}

\begin{figure*}[t]
%\begin{figure}[t]
\centerline{\includegraphics[width=0.8\textwidth]{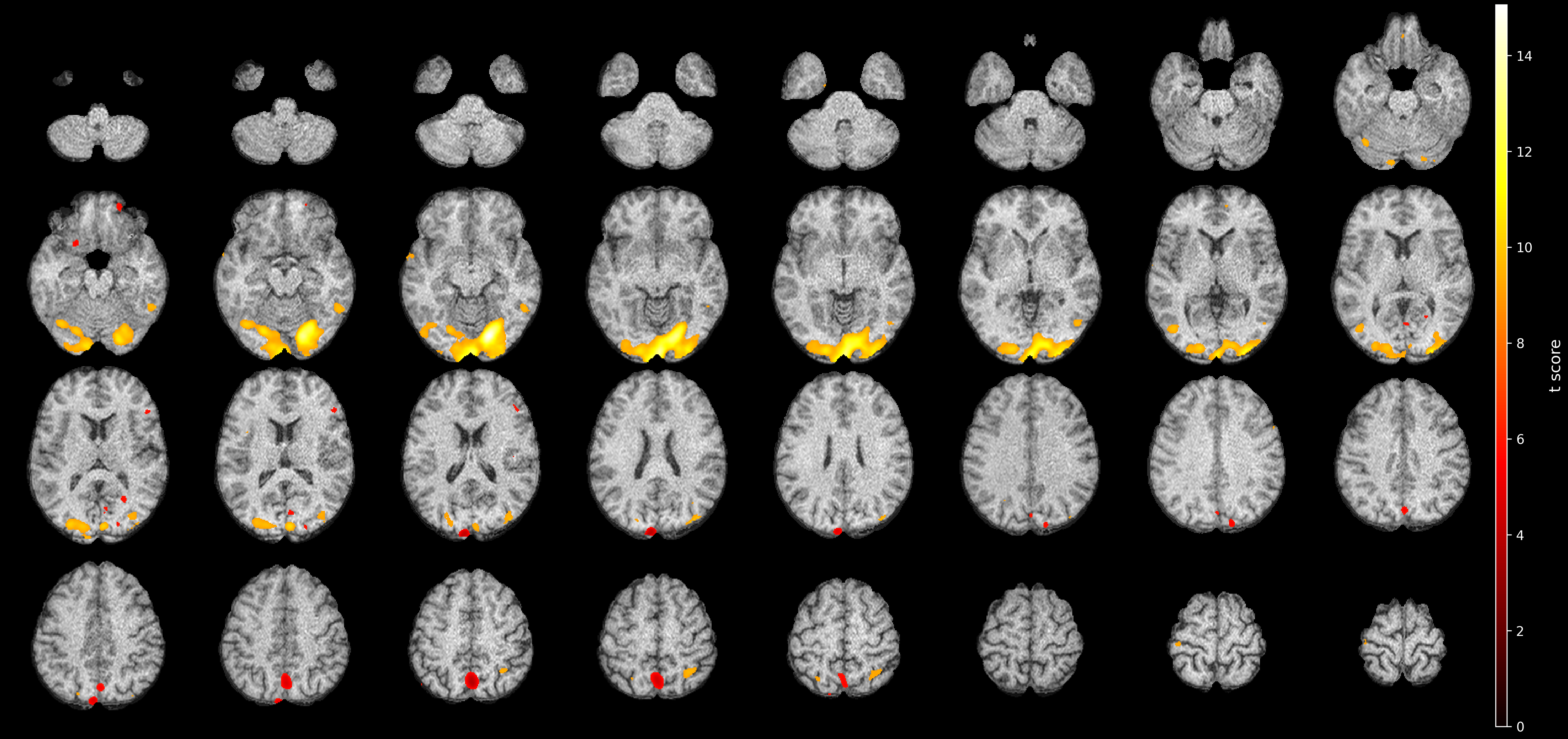}}
\caption{Visual activation and statistical reliability at 0.55T. Whole-brain axial montage of t-statistic maps from a representative subject, generated from 10 minutes of concatenated task data. Activation maps are overlaid on the 0.55T T1-weighted anatomical image. Significant, contiguous clusters are localized to the primary visual cortex (V1) and surrounding occipital regions. Statistical maps are thresholded at $q < 0.05$ (FDR-corrected) to demonstrate the spatial extent and reliability of the BOLD signal at 0.55T.
%Statistical maps and activation reliability for the visual experiment. The 10-minute concatenated data yields focused engagement of V1 with metrics comparable to 3T systems. 
%\color{blue}{(Note: This figure is a template showing fMRI t-statistic activation maps and will be replaced with the final results from the subjects included in this draft. The final version will show either whole-brain slices, as in the current figure, or selected axial slices targeting the region of interest in this participant.)}
}\label{fig3}
\end{figure*}
%\end{figure}

\subsection{Processing Pipelines}
The fMRI data were analyzed using the BrainSuite fMRI Pipeline (BFP) and SPM12 \cite{b22, b23, b24}. Preprocessing included slice-timing correction, motion correction, 8mm FWHM spatial smoothing, and 0.005 Hz high-pass filtering \cite{b24}. To achieve robust statistical power, two 5-minute runs were concatenated to form a 10-minute dataset for the final analysis \cite{b17}. 

The EEG denoising pipeline used the BrainVision Analyzer \cite{b25} and Brainstorm \cite{b26} to address the specific noise profile of the 0.55T scanner. GA correction was performed using Average Artifact Subtraction (AAS) \cite{b19}. BCG artifacts were removed using a pulse-artifact subtraction method based on the concurrent ECG channel. BCG amplitudes were measured on R-peak-locked epochs ($-$100 to $+$600~ms) from the scanner-OFF condition using the global field power (GFP) temporal standard deviation across EEG channels, the metric of Debener et al.\ \cite{debener2008}, after 1--40~Hz bandpass filtering and 5--95\% trimming across cycles. Channels with persistent abnormal impedance or amplitude drift were first excluded based on visual inspection, and the remaining channels were re-referenced. Independent Component Analysis (ICA) was then applied to isolate residual ocular and scanner-related artifacts from neural components. Ocular artifact components were identified based on their spatial topography and high temporal correlation with the Fp1 and Fp2 channels containing eye-blink activity. These components were removed, and the signal was reconstructed from the retained components. A final visual inspection of the reconstructed time series was performed to confirm signal quality. Representative EEG traces across all three recording conditions and after artifact subtraction are shown in Fig.~\ref{fig2}.

%\added[id=P]{BCG amplitudes were measured on R-peak-locked epochs ($-$100 to $+$600~ms) from the scanner-OFF condition using peak-to-peak on Oz (1--40~Hz bandpass, 5--95\% trimmed mean) and global field power (GFP) temporal standard deviation across EEG channels \cite{debener2008}.} 

%\replaced[id=P]{Channels with persistent abnormal impedance or amplitude drift were excluded based on visual inspection, and the remaining channels were re-referenced. Independent Component Analysis (ICA) was then applied to isolate residual artifacts from neural activity. Components were identified as artifacts based on their spatial topography and temporal correlation with known noise sources, including ocular activity and scanner-related interference. Ocular artifact components were identified based on their frontal topography and high temporal correlation with the Fp1 and Fp2 channels containing eye-blink activity. These components were removed, and the signal was reconstructed from the retained components. A final visual inspection of the reconstructed time series was performed to confirm signal quality.}{Finally, Independent Component Analysis (ICA) was applied to isolate residual artifacts from neural components, followed by visual inspection of the final data, where a few channels were marked as bad and excluded from the analysis.}

\subsection{Multimodal Data Integration}
To evaluate the coupling between electrophysiological activity and the hemodynamic response, we performed a voxel-wise correlation analysis comparing the observed BOLD fMRI signal with two distinct predictive models: a data-driven EEG predictor and a hypothesis-driven block-design model.

% \begin{figure*}[t]
% %\begin{figure}[t]
% \centerline{\includegraphics[width=0.8\textwidth]{3.png}}
% \caption{Visual activation and statistical reliability at 0.55T. Whole-brain axial montage of t-statistic maps from a representative subject, generated from 10 minutes of concatenated task data. Activation maps are overlaid on the 0.55T T1-weighted anatomical image. Significant, contiguous clusters are localized to the primary visual cortex (V1) and surrounding occipital regions. Statistical maps are thresholded at $q < 0.05$ (FDR-corrected) to demonstrate the spatial extent and reliability of the BOLD signal at 0.55T.
% %Statistical maps and activation reliability for the visual experiment. The 10-minute concatenated data yields focused engagement of V1 with metrics comparable to 3T systems. 
% %\color{blue}{(Note: This figure is a template showing fMRI t-statistic activation maps and will be replaced with the final results from the subjects included in this draft. The final version will show either whole-brain slices, as in the current figure, or selected axial slices targeting the region of interest in this participant.)}
% }\label{fig3}
% \end{figure*}
% %\end{figure}

\begin{figure}[t]
%\centerline{\includegraphics[width=0.8\columnwidth]{4.png}}
\centerline{\includegraphics[width=0.90\columnwidth]{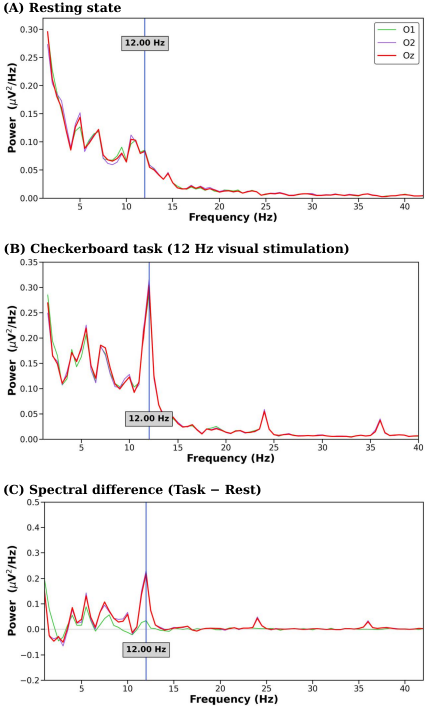}}
\caption{Spectral characterization and task-evoked oscillatory responses at 0.55T. Power Spectral Density (PSD) analysis across recording conditions (Rest vs. Checkerboard) for occipital channels ($O_z$, $O_1$, and $O_2$) from a representative subject demonstrates the fidelity of neural signal recovery. (A) The spontaneous alpha rhythm (8--13~Hz) is preserved in the resting-state condition following artifact subtraction, serving as a baseline indicator of signal integrity. (B) During the visual task, the power spectrum exhibits a robust fundamental response at the 12~Hz stimulation frequency, with distinctly identifiable higher-order harmonics at 24~Hz and 36~Hz. (C) The spectral difference plot highlights the narrow-band power localized to these frequencies, confirming that the denoising pipeline effectively suppresses scanner-induced interference while retaining both physiological and task-specific spectral features.}
\label{fig4}
\end{figure}

\subsubsection{Construction of the EEG-Informed Predictor}
The \textit{expected BOLD} time course was derived directly from the subject’s neural data by isolating the Steady-State Visual Evoked Potential (SSVEP) \cite{b2} from the Oz electrode, which is maximally sensitive to visual stimulation. The preprocessing pipeline comprised three steps: 1) \textbf{spectral isolation}, where the preprocessed Oz time series was bandpass filtered between 11 Hz and 13 Hz (20th-order IIR filter) to isolate the target checkerboard frequency; 2) \textbf{power extraction}, in which the instantaneous amplitude envelope was computed using the Hilbert transform to quantify neural power fluctuations over time; and 3) \textbf{hemodynamic modeling}, where this power envelope was downsampled to the fMRI repetition time (TR = 3.0~s) and convolved with a canonical Hemodynamic Response Function (SPM-HRF) to account for neurovascular delay. This procedure converts the electrophysiological signal into a hemodynamic predictor that can be directly compared with the BOLD time course.

\subsubsection{Statistical Mapping}
We calculated the Pearson correlation coefficient ($r$) between the measured fMRI time series at each voxel and two predictors: (1) the standard Block Design boxcar model convolved with the canonical HRF, and (2) the EEG-derived power envelope. This generated two independent spatial maps representing voxels synchronized with the experimental timing and voxels synchronized with the subject's actual neural power at 12 Hz.

%The EEG denoising pipeline utilized EEGLAB, FMRIB, and Brainstorm to address the specific noise profile of the 0.55T scanner \cite{b23, b24, b25}. GA correction was performed using Average Artifact Subtraction (AAS). BCG artifacts were removed using a pulse-artifact subtraction method based on the concurrent ECG channel. Finally, Independent Component Analysis (ICA) was applied to isolate residual artifacts from neural components.

\section{Results}

\subsection{Significant Visual Activation at 0.55T}
% Significant activation ($p < 0.05$, FDR corrected) was consistently observed in the visual cortex across participants, as shown in the statistical maps in Fig. \ref{fig3}. Utilizing the 10-minute concatenated protocol, activation was localized to the primary visual cortex (V1) and adjacent visual areas. 
Significant activation ($p < 0.05$, FDR corrected) was consistently observed in the visual cortex across subjects, with a representative whole-brain t-statistic map shown in Fig.~\ref{fig3}. Using the 10-minute concatenated protocol, activation was localized to the primary visual cortex (V1) and adjacent visual areas in both subjects. The analysis yielded a mean t-score of 9.0, demonstrating that detection of block-design BOLD signal changes is feasible at 0.55T with extended scanning durations \cite{b17}.

\subsection{EEG Signal Quality and Artifact Characterization}

Analysis of the 0.55T environment reveals a distinct and favorable artifact profile. Quantitatively, the GFP temporal standard deviation of the BCG \cite{debener2008}, measured in the scanner-OFF condition, was 5.8~$\pm$~2.2~$\mu$V at 0.55T (N=2 subjects), falling on the linear regression through Debener's published values at 1.5, 3, and 7~T (15.3, 24.9, 51.5~$\mu$V), providing the first empirical extension of the BCG field-strength scaling law to mid-field MRI. This reduction facilitated effective denoising using standard gradient and BCG subtraction techniques without the need for aggressive filtering that might compromise neural data. Power Spectral Density (PSD) analysis of the cleaned data (Fig. \ref{fig4}) confirms the preservation of the spontaneous alpha rhythm (8--13~Hz) in the resting-state condition and the recovery of the 12~Hz SSVEP at the stimulation frequency and its harmonics in the task condition.

\begin{figure}[t]
%\centerline{\includegraphics[width=\columnwidth]{5.png}}
\centerline{\includegraphics[width=\columnwidth]{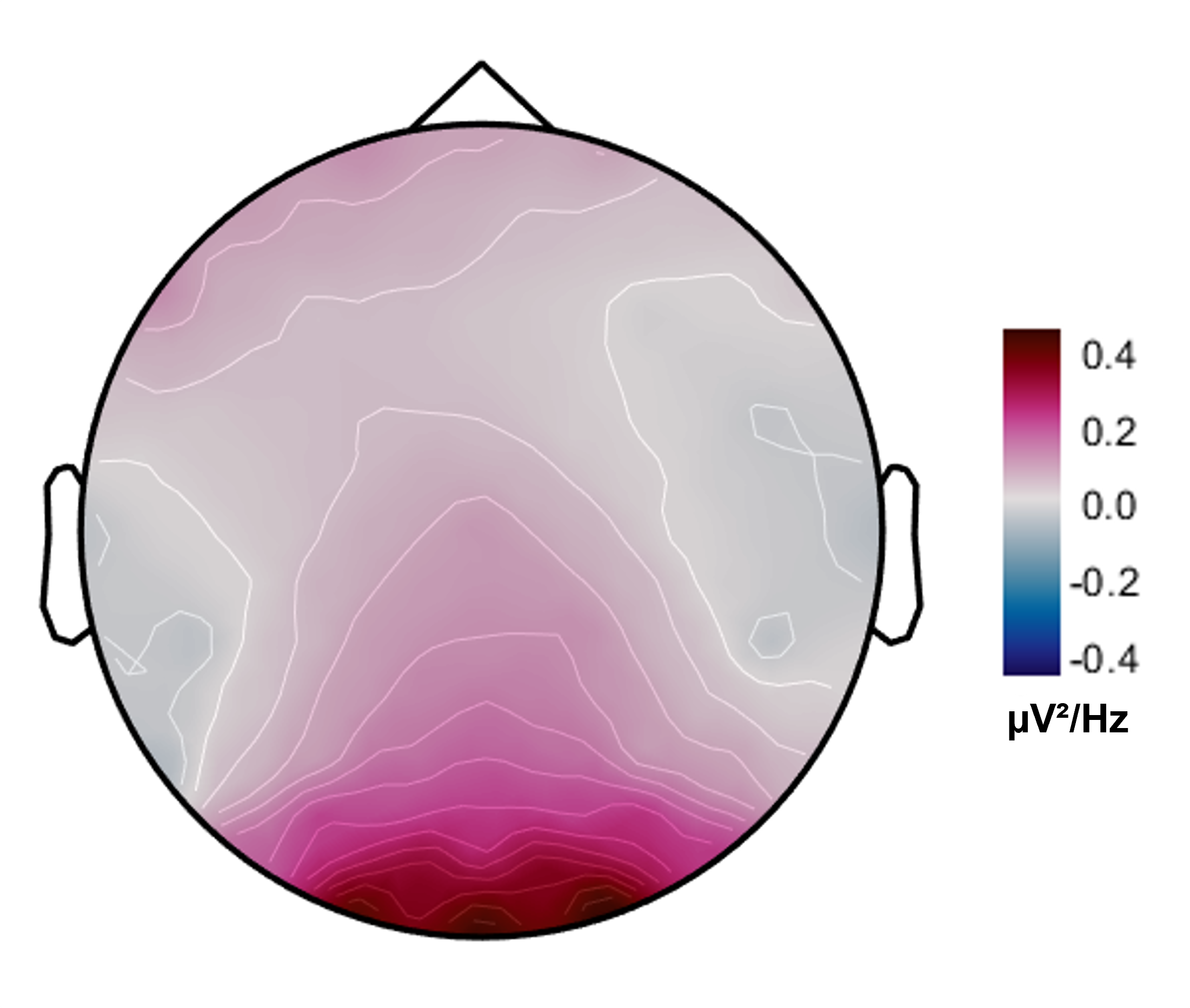}}
\caption{Topographical distribution of SSVEP power modulation at 0.55T from a representative subject. The heatmap illustrates the spatial distribution of the spectral power difference (Checkerboard $-$ Rest) at the 12~Hz stimulation frequency. A highly focal activation is centered over the occipital electrodes ($O_z$, $O_1$, and $O_2$), consistent with the primary generators of the visual evoked response. This localized topography confirms that the signal recovery pipeline suppressed widespread scanner-induced artifacts while preserving the underlying spatial integrity of the neural data in the mid-field environment.}
\label{fig5}
\end{figure}

Furthermore, the topographical distribution of the SSVEP power modulation (Fig. \ref{fig5}) indicates focal activation over occipital electrodes, demonstrating that spatial information is preserved after artifact removal.

%\begin{figure}[t]
%\centerline{\includegraphics[width=\columnwidth]{5.png}}
%\caption{Topographical distribution of SSVEP power modulation. The heatmap reveals a focal activation centered over the occipital electrodes (Oz, O1, O2) during the task. (or The heatmap displays the magnitude of the 12 Hz power increase (Task - Rest), revealing a highly focal activation centered over the occipital electrodes (Oz, O1, O2).) \color{blue}{(Note: This figure will be updated to show topographic plots comparing the Rest and Checkerboard conditions, as well as their difference, across three acquisition settings (outside the scanner, scanner off, and scanner on). Additional details on data quality control will be added to the main text, including the definition of “good” channels, the identification and removal of bad channels based on visual inspection, and an appropriate reference citation. Mention the difference.)}}\label{fig5}
%\end{figure}

\subsection{Spatial Correspondence of Activation Maps}
%The multimodal correlation analysis demonstrated that the EEG-informed model generates a similar level of activity in the occipital lobe as the standard Block Design. As shown in Fig. \ref{fig6}, both the EEG-informed predictor (Fig. \ref{fig6}A) and the Block Design model (Fig. \ref{fig6}B) identified significant activation clusters localized to the primary visual cortex (V1) and adjacent visual areas. This localization is consistent with the expected generator of the SSVEP signal, confirming that the EEG envelope is a robust predictor of BOLD variance in the visual cortex.

The multimodal correlation analysis showed a similar spatial pattern between the data-driven EEG predictor and the standard block design in the occipital lobe. As shown in Fig. \ref{fig6}, the EEG-informed model (Fig. \ref{fig6}A), based on the $O_z$ power envelope, identified significant activation clusters localized to the primary visual cortex (V1) and surrounding visual areas. This spatial map is similar to the standard block-design activation (Fig. \ref{fig6}B), suggesting that the recovered EEG signal captures the expected physiological changes in the visual cortex. These results provide initial evidence that at 0.55T, the $12$~Hz rhythmic power can be used to predict the task-related fMRI BOLD response.

\section{Discussion and Conclusion}
This pilot study suggests that simultaneous EEG-fMRI is feasible and promising at 0.55T. By employing a 10-minute concatenated acquisition protocol, we observed significant BOLD activation in the visual cortex, comparable to patterns typically reported in conventional-field studies, despite the inherent SNR challenges of 0.55T \cite{b17}. These findings support the utility of high-performance 0.55T systems for functional neuroimaging when acquisition protocols are tailored to the specific characteristics of the lower-field environment.

A notable advantage observed in this environment is the favorable artifact profile for EEG. The magnitude of the BCG artifact appeared reduced compared to high-field magnets, consistent with the physics of magnetic field interactions. This reduction likely contributed to the efficacy of standard denoising pipelines, allowing for the preservation of the spontaneous alpha rhythm and the SSVEP without the need for aggressive filtering that might compromise data integrity. These findings suggest that simpler and less computationally intensive denoising pipelines may be sufficient at 0.55T, which is particularly relevant for real-time and closed-loop EEG-fMRI applications.  Closed-loop neurofeedback is an established application of simultaneous EEG-fMRI \cite{ciccarelli2023, lioi2020}, in which electrophysiological and hemodynamic signals are monitored in real time to guide self-regulation paradigms, and the reduced denoising demand at 0.55~T may broaden its practical accessibility. Beyond neurofeedback, real-time multimodal acquisitions that combine MRI with electrophysiological recordings are emerging as a platform for studying complex behaviors such as speech production \cite{lee2026arxiv}. The reduced artifact burden at 0.55~T may also facilitate investigation of higher-frequency neural activity, including gamma-band activity, that is more difficult to recover at high field strengths. Future work will evaluate these directions in larger cohorts and real-time acquisition settings.

%\begin{figure}[t]
%\centerline{\includegraphics[width=\columnwidth]{6.png}}
%\centerline{\includegraphics[width=\columnwidth,height=0.33\textheight,keepaspectratio]{6.png}}
%\includegraphics[width=\columnwidth,height=0.35\textheight,keepaspectratio]{6.png}
%\caption{Multimodal activation maps using correlation analysis. Representative correlation surface maps from a single participant display the spatial distribution of Pearson correlation coefficients (r). (A) Activation map derived using the Oz EEG signal (12 Hz envelope) as the predictor. (B) Activation map derived using the standard Block Design (Boxcar + HRF). The spatial overlap confirms the EEG envelope is a robust predictor of BOLD variance.  \color{blue}{(Note: This figure will be updated to include t-statistic maps overlaid on the subject’s T1-weighted image. The values will be refined in the next version of the figure. If a surface-based map is shown, the colormap will be changed to a gray cortical surface with curvature, similar to the FreeSurfer surface plot. I’ll add a few equations and an explanation of how the EEG regressor and voxel-wise correlation are computed, and I’ll also reference related prior work)}}\label{fig6}
%\end{figure}

\begin{figure}[t]
\centerline{\includegraphics[width=\columnwidth]{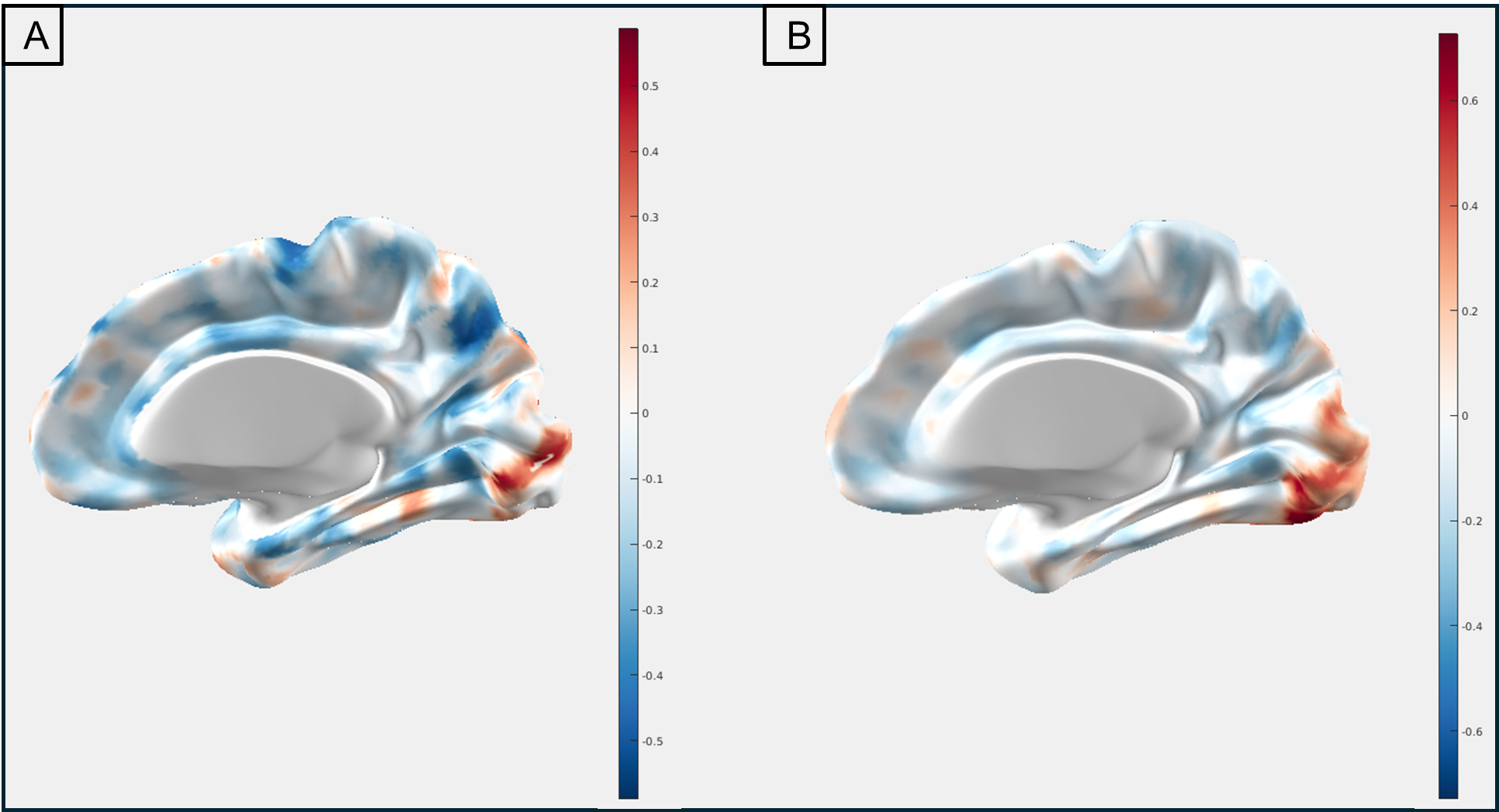}}
\caption{Multimodal functional mapping and neurovascular coupling at 0.55~T. Cortical correlation maps from a 5-minute run in a representative subject demonstrate the spatial distribution of Pearson correlation coefficients ($r$). (A) Functional map derived from the voxel-wise correlation between the measured BOLD signal and the data-driven 12~Hz power envelope extracted from the $O_z$ electrode. (B) Functional map derived using the hypothesis-driven block design (Boxcar convolved with a canonical HRF). The high degree of spatial correspondence between the EEG-informed predictor and the standard model validates the integrity of the multimodal integration and confirms that the BOLD signal at 0.55T remains tightly coupled to underlying electrophysiological power modulation.}
\label{fig6}
\end{figure}

The multimodal analysis provides preliminary evidence for neurovascular coupling at 0.55T. The spatial correspondence between the activation maps derived from the standard block design and the data-driven EEG regressor (Fig. \ref{fig6}) indicates that the BOLD signal in the visual cortex is tightly coupled to the underlying electrophysiological power in the 12 Hz band. This result serves as a technical validation of the multimodal acquisition pipeline, suggesting that the EEG signal can serve as a viable predictor of hemodynamic activity in this environment. Neurovascular coupling is a fundamental mechanism underlying the BOLD signal \cite{girouard2006neurovascular}, and multimodal EEG-fMRI frameworks are increasingly investigated for objective characterization of neurological and neurodevelopmental conditions \cite{b1, ranjbar2025beyond, torabi2026neuromamballm}.

These results, drawn from a pilot cohort (N = 2), should be interpreted as preliminary and require validation in larger cohorts. Nevertheless, they provide proof-of-concept evidence for multimodal imaging at 0.55T. The high-field comparison values reported here are drawn from the published literature (Debener et al.\ \cite{debener2008}). A within-subject 0.55T vs.\ high-field acquisition is an important next step to confirm the BCG reduction. The demonstration of significant BOLD activation, high-fidelity recovery of focal SSVEP signatures, and the observed spatial correspondence between electrophysiological and hemodynamic signals, combined with the benefits of accessible hardware and reduced artifact interference, collectively positions high-performance 0.55T MRI as a potentially advantageous platform for future clinical and research applications, particularly where high-field scanners may be unsuitable or unavailable \cite{b15}.

\section*{Ethics Statement}
The experimental procedures involving human subjects described in this paper were approved by the Institutional Review Board of the University of Southern California.

%\section*{Acknowledgment}
%The authors acknowledge support from the National Institutes of Health, National Science Foundation, and Siemens Healthineers.  We thank Sophia X. Cui for technical support, and Mary Yung for research coordination.

\section*{Acknowledgment}
%The authors acknowledge support from the National Institutes of Health, National Science Foundation, and Siemens Healthineers. 
We thank Sophia X. Cui (Siemens Healthineers) for technical support, and Mary Yung for research coordination.

\end{document}